\documentstyle[pre,aps,multicol,epsf]{revtex}
\tighten

\begin{document} 
\draft

\title{
The Potts Fully Frustrated model: 
Thermodynamics, percolation and dynamics 
in 2 dimensions
}  
\author{
Giancarlo Franzese \thanks{Present address: Center for Polymer Studies,
Boston University, 590 Commonwealth Avenue, Boston, MA 02215}
}
\address{
Dip. di Fisica ``E.Amaldi'', Universit\`a Roma Tre, 
via della Vasca Navale 84, I-00146 Roma, Italy\\
Istituto Nazionale per la Fisica della Materia - unit\`a di Napoli 
Mostra d'Oltremare Pad.19 I-80125 Napoli, Italy
}

\date{\today}

\maketitle 

\begin{abstract} 
We consider a Potts model diluted by fully frustrated Ising spins. 
The model corresponds to a fully frustrated Potts model  
with variables having an integer absolute value and a sign. 
This model presents precursor phenomena of a glass transition
in the high-temperature region. 
We show that the onset of these phenomena  can be related to a
thermodynamic transition. 
Furthermore this transition can be mapped onto a percolation transition.
We numerically study the phase diagram in 2 dimensions (2D) for this
model with frustration and {\em without} disorder and we compare it 
to the phase diagram of $i)$ the model with frustration {\em and} disorder  
and of $ii)$ the ferromagnetic model. 
Introducing a parameter that connects the three models, we generalize
the exact expression of the ferromagnetic Potts transition temperature
in 2D to the other cases.
Finally, we estimate the dynamic critical exponents related to the
Potts order parameter and to the energy.

\end{abstract} 



\begin{multicols}{2}
\narrowtext
\section{Introduction}

The glass transition temperature $T_g$ for liquids 
is defined experimentally as the
onset of calorimetric anomalies \cite{calorimetric}. 
It is usually understood that
$T_g$ is not related to a thermodynamic transition \cite{Odagaki},
but to the slowing down of one or more degrees of freedom. This slowing
down prevents the system to reach the equilibrium
\cite{Angell,Gotze}. 
This question is of interest because 
almost all liquids can form glasses if cooled at high
enough rate. Moreover, many other materials as for example 
polymers, microemulsions, granular material, vortex
glasses, ionic conductors, colloids, plastic glassy crystals and spin glasses
(SGs) \cite{SG} show glassy properties.

Even well above $T_g$, where glassy systems actually can
equilibrate, they  
show experimentally dynamic anomalies as precursor phenomena of the glass
transition \cite{Angell,glass,Andreozzi,Feldman}.
From a theoretical point of view, one of the open questions is if 
these precursor phenomena are related to the thermodynamics of the system
\cite{Odagaki,Gotze,Randeria,vari}, or if they are not, like the glass
transition occurring at lower temperature $T_g$.
In particular, for the Ising SG \cite{SG} 
such a relation between precursor phenomena
and a thermodynamic free energy {\em essential} singularity has been shown 
\cite{Randeria,Griffiths,Ogielski}. 
Indeed, in this model there is a dynamic anomaly at a temperature $T^*$.
Above $T^*$ the relaxation processes have an exponential behavior, while
below $T^*$ they have a non-exponential behavior. Theoretical and
numerical evidences show that $T^*$ coincides with the Griffiths
temperature $T_c$ \cite{Griffiths}. This $T_c$ is the transition
temperature that the model would have if the frustration due to disorder
is removed. A way to remove the frustration is, 
for example, to substitute every antiferromagnetic interaction with a
ferromagnetic interaction.
In general, removing the frustration, the model will have 
ferromagnetic regions and antiferromagnetic regions and $T_c$ will be  
the transition temperature of the
unfrustrated model.
To be more precise, the free energy of the Ising SG in external field
has a singularity which disappears in the limit of zero external field
and which occurs at a temperature that goes to $T_c$ in the same limit.
This singularity is present only in disordered systems.
The relation $T^*=T_c$ has also been shown to be valid numerically in 2
dimensions (2D) for a model with Potts variables that generalizes  
the Ising SG \cite{breve}. This generalized model will be considered
in the following and we will refer to it as the Potts SG.

Until now we talked about glass dynamics in systems with disorder, but
numerical simulations show that it is possible to observe glassy
behavior with precursor phenomena for spin systems {\em without}
disorder but {\em with} frustration \cite{FFdCC,KL}.  These systems, due to
the lack of disorder, are 
more suitable for a theoretical approach \cite{pezzella,dLP}. 
In particular, one can try to
answer the  question about the relation between the precursor
phenomena and the thermodynamics of the system.	

In Ref.\cite{FFdCC} it
was considered a simple case: The fully frustrated (FF) Ising model
\cite{Villain} 
where ferromagnetic and antiferromagnetic interactions are ordered in
such a way that any lattice cell has an odd number of antiferromagnetic
interactions (i.e. is frustrated).  It is shown (by simulations in 2D
and 3D) that the onset $T^*$ of non-exponential relaxation processes is
related to a random-bond percolation transition \cite{FFdCC}.  As for
the Ising SG, it is possible to generalize the FF Ising model to a FF
model with Potts variables \cite{breve}: The Potts FF  model that
will be described in details in the following. In Ref.\cite{breve} the
dynamics of the Potts FF model was compared with the dynamics of the Potts SG
model and some anticipations on the relation between precursor phenomena
and thermodynamics were given. In this work we study in details the
thermodynamics of the Potts FF model, showing  that $T^*$ corresponds to the
thermodynamic transition temperature $T_p$ of the Potts
variables. It is important to
note that in any FF model the Griffiths temperature $T_c$ cannot be
defined for the lack of disorder. Therefore $T_c$ cannot play any role in
these cases.

The comparison of the results presented here with the analogous study of
the Potts SG model with disorder \cite{PSG} allows to give insight on the
role of disorder.  To this goal we introduce a formal parameter $0\leq X
\leq 1$  that connects both models. 
For $X=0$ we have the ferromagnetic Potts model \cite{Wu} (without
frustration and without disorder).
For $0<X<1$  we have  the disordered {\em and} frustrated Potts model.
In particular, for $X=0.5$ we have the Potts SG model.
For $X=1$ we have the Potts FF model (without disorder). 

From Ref.\cite{PSG} and from the present work, it is possible to show that
for $X=0.5$ and $X=1$ there are two thermodynamics transitions.
The lower transition 
is an Ising SG or a FF Ising transition, respectively. 
The upper transition at $T_p$ is in the universality class of a
ferromagnetic Potts transition.
Furthermore we show that $T_p$ corresponds to a percolation temperature. 
Moreover, we show how it is
possible to generalize the exact expression of $T_p$ for the model with
$X=0$ in 2D \cite{Wu} to the cases $X=0.5$ and $X=1$.

The organization  of the paper is the following.
In Sec.~II we introduce the model and the known results 
for $X=0$ (ferromagnetic case),
for $X=0.5$ (disordered and frustrated case)
and for $X=1$ (ordered and frustrated case).
In Sec.~III
we introduce  the cluster
formalism used to map the upper thermodynamic transition at $T_p$ 
onto a percolation transition.
In Sec.~IV we present the phase diagram in
2D for  $X=1$ as result of Monte Carlo (MC) simulations
and we compare it with the cases $X=0$ and $X=0.5$.
In Sec.~V 
we use the spin-flip MC dynamics to study the dynamic critical
exponent and the temperature $T^*$, onset of stretched exponentials.
In Sec.~VI we give the summary and conclusions.

\section{The Model}

Structural glasses, such as  dense molecular glasses, plastic
crystal, or orto-therphenyl at low
temperature, can be modeled to first approximation as
systems with orientational degrees of freedom frustrated by geometrical
hindrance between non-spherical molecules. 
For this reason  we
will consider the lattice model introduced in Ref.~\cite{CdLMP}, 
where the orientational degrees of freedom are represented by 
Potts variables \cite{Wu} with $s$ states
($\sigma_i=1,\dots,s$) and the frustration is modeled by means of 
ferro/antiferromagnetically interacting Ising spins
($S_i=\pm1$), coupled to the Potts variables.

The model is defined by the Hamiltonian 
\begin{equation}
H_s\{S_i,\sigma_i,\epsilon_{i,j}\}
=-sJ\sum_{\langle i,j \rangle} \delta_{\sigma_i, \sigma_j}
(\epsilon_{i,j}S_i S_j+1)
\label{hamiltonian}
\end{equation}
where the sum is extended over all the nearest neighbor  sites,
$J$ is the strength of interaction,
$\epsilon_{i,j}=\pm 1$ is a quenched variable that represents the sign
of the ferro/antiferromagnetic interaction 
and $\delta_{n,m}=0,1$ is a Kronecker delta. 
To emphasize that the Ising and the Potts variables are interdependent,
we can rewrite the Hamiltonian in Eq.(\ref{hamiltonian}) as
\begin{equation}
H_s\{\tau_i,\epsilon_{i,j}\}=
-2sJ\sum_{\langle i,j \rangle} \delta_{\epsilon_{i,j}\tau_i, \tau_j}
\label{hamiltonian2}
\end{equation}
where $\tau_i\equiv S_i\sigma_i=\pm 1, \pm 2, \dots , \pm s$ is a
variables with $2s$
states and the frustration now is explicitly on the new variable.
Any $\tau_i$ has an absolute value $\sigma_i$ and a sign $S_i$.

\begin{figure}
\mbox{ \epsfxsize=8cm \epsffile{ 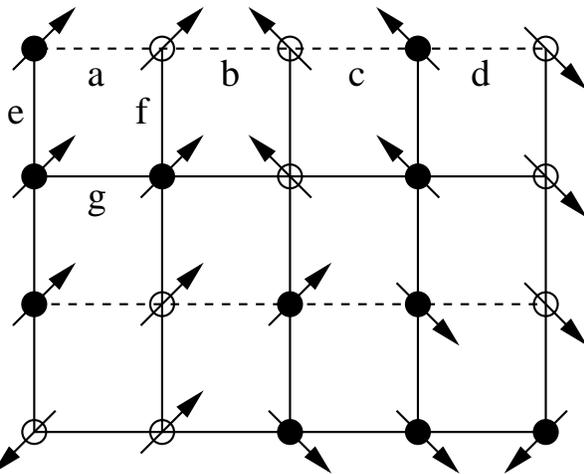 } }
\caption{
Example of Potts fully frustrated  model on a square lattice: 
on each vertex there is variable $\tau_i=S_i\sigma_i=\pm 1, \pm 2 \dots \pm s$ with
$s=4$ in the figure. Here we represent the sign ($S_i$) of each $\tau_i$ by an
open or a full dot (respectively positive and negative, for example)  
and its orientational state ($\sigma_i$) by an arrow pointing in 
4 different directions. 
Ferromagnetic (antiferromagnetic) interactions are represented by 
full (dotted) lines.
}
\label{lattice}
\end{figure}

The model depends on the interaction configuration $\{\epsilon_{i,j}\}$.
Possible choices are the following.  ($i$)  If all $\epsilon_{i,j}=1$
(i.e. all interactions are ferromagnetic), the Eq.(\ref{hamiltonian2})
is the Hamiltonian of the ferromagnetic Potts model \cite{Wu} with
variables $\tau_i$ with an even ($2s$) number of states.  It shows a
thermodynamic transition at $T_c(s)$ whose order depends on $s$.  ($ii$)
If $\epsilon_{i,j}$ are quenched random variables, the model corresponds
to the  Potts SG.  This model  is a generalization of the Ising SG model
that is recovered for $s=1$.  It shows two thermodynamic transitions
\cite{PSG}.  The lower is a SG transition at $T_{SG}(s)$.  The upper is
a Potts transition at $T_p(s)>T_{SG}(s)$.  The transition at $T_p(s)$ is
in the universality class of a ferromagnetic $s$-states Potts model.
Another relevant temperature for this model is the  temperature $T_c(s)$
defined for the previous case (the ferromagnetic $2s$-states Potts
model).  
Indeed, it is possible to show that for finite external field a free
energy (Griffiths) singularity arises \cite{Griffiths}. 
In the limit of external field going to zero, the temperature at which
this singularity occurs goes to $T_c(s)$ and the singularity vanishes.
$T_c(s)$ is the Griffiths temperature for this model.  
Furthermore, numerical simulations \cite{breve} show that
$T_c(s)>T_p(s)$ corresponds to the onset $T^*(s)$ of
non-exponential correlation functions for the Ising spins $S_i$.  This
result generalizes what happens in the Ising SG ($s=1$ case), where at
the Griffiths temperature $T_c(1)$ non-exponential correlation
function are seen \cite{Ogielski}.  It is worth to note that the Ising
spins are critical  at $T_{SG}(s)$, that is  well below $T_c(s)=T^*(s)$.
Moreover, note that the relevant Griffiths temperature for this model is
$T_c(s)$ of the variables $\tau_i=S_i\sigma_i$ and not the Griffiths
temperature of the variables $S_i$ alone, that in our notation is $T_c(1)$.
($iii$)  If there is an odd number of $\epsilon_{i,j}=-1$ for each
elementary cell (as in Fig.~\ref{lattice}), the system is fully
frustrated (FF).  It means that at least one interaction per cell is not
satisfied, i.e. the relative energy contribution is $0$ instead of
$-2sJ$ (as for the edges $b$, $d$ and $f$ in Fig.~\ref{lattice}).  The
model is called Potts FF and is a generalization of the FF Ising models
\cite{Villain} that is recovered for $s=1$.  It has frustration but no
disorder.  For any integer $s\geq 1$ the model has in 2D a second-order
phase transition at $T=0$.  Non-exponential correlation functions are
reported below a finite temperature $T^*(s)$ for $s=2$, 1 and $1/2$
(the latter case is defined in Sec.III) \cite{FFdCC,breve}.  This
dynamic anomaly cannot be related, as in the previous case, to the
Griffiths temperature.  Indeed, the Griffiths temperature is not defined
in FF models for the lack of disorder.  The aim of this work is to study
the phase diagram of the FF model as function of $s$ and to show that
$T^*(s)$ corresponds to a Potts transition.  Furthermore, using a
percolation approach, it is possible to show that the Potts transition
corresponds to a percolation transition, defined also for non-integer
values of $s$.

The previous three cases can be generalized in a class of Hamiltonians.
Consider, as model parameter, the density $X$ of frustrated elementary
cells.  For $X=0$ there are no frustrated cells (like in a ferromagnet),
while for $X=1$ every cell is frustrated (FF case).  For $0<X<1$ it is
possible to partition the lattice in two non-overlapping sub-sets $U(X)$
and $F(X)$ of unfrustrated cells and frustrated cells, respectively.
Therefore the Eq.(\ref{hamiltonian2}) can be written as
\begin{eqnarray}
\label{hamiltonianX}
\nonumber 
H_{s,X}\{\tau_i\}\equiv H_s\{\tau_i,\epsilon_{i,j}(X)\}= -2sJ\sum_{\langle
i,j \rangle\in U(X)} \delta_{\tau_i, \tau_j} \\ 
 -2sJ\sum_{\langle i,j
\rangle\in F(X)} \delta_{\epsilon_{i,j}\tau_i, \tau_j} \ .
\end{eqnarray}
In this way the ferromagnetic Potts model with an even number of states
(case $i$) is recovered for $X=0$, since $F(X=0)$ is empty;  The Potts
SG model (case $ii$) for $X=0.5$;  The Potts FF model (case $iii$) for
$X=1$, since $U(X=1)$ is empty;  Intermediate disordered models are
obtained for other values of $0<X<1$.

To make the comparison between the $X=0.5$ and $X=1$ model, we will
follow the line of Ref.~\cite{PSG} where the phase diagram of the former
has been studied.  In particular, for $X=1$ we consider a square lattice
with  one $\epsilon_{i,j}=-1$ per cell as shown in Fig.\ref{lattice}.

\section{Percolation}

We now introduce the percolation map following the Ref.~\cite{CdLMP}.
It is a generalization to $X\geq 0$ of Fortuin-Kasteleyn \cite{FK-CK}
clusters formalism, defined originally only for  $X=0$.  
In Ref.~\cite{CdLMP} it is shown that the partition function $Z_{s,X}$ of the
Hamiltonian in Eq.(\ref{hamiltonianX}) can be written in terms of bond
configurations $C$
\begin{equation}
Z_{s,X}\equiv \sum_{\{\tau_i\}}e^{-H_{s,X}\{\tau_i\}/(k_B T)}=\sum_C W_{s,X}(C)
\label{partition_function}
\end{equation}
where $k_B$ is the Boltzmann constant,  $W_{s,X}(C)=0$ if $C$ includes any
{\em frustrated loop} (defined below), otherwise
\begin{equation}
W_{s,X}(C)=p^{|C|}(1-p)^{|A|}(2s)^{N(C)}
\label{W_q}
\end{equation}
where $p=1-\exp[-2sJ/(k_B T)]$  is the probability to place a bond
between two nearest neighbor sites, $N(C)$ is the number of clusters
(defined as maximal sets of connected bonds) in the configuration $C$,
$|C|$ is the number of bonds and $|C|+|A|$ is the total number of
interactions.  A loop of bonds is called {\em frustrated} if the product
of all the signs $\epsilon_{i,j}$ of the interactions along it is
equal to $-1$.  For a frustrated loop there is no $\{\tau_i\}$ configuration
able to minimize the energy of all the interactions along it.  
An example of such a frustrated loop is the one composed by the edges
$a$, $e$, $g$, $f$ in Fig.\ref{lattice} or a loop composed by the
external edges of any three adjacent elementary cells in a FF lattice (or
of any odd number of adjacent cells).

Note that for $X=0$ there are no frustrated loops and the original
Fortuin-Kasteleyn cluster definition is recovered.  
Since the above cluster definition holds for any $s$ and $X$, 
for any model described by the general Hamiltonian in
Eq.(\ref{hamiltonianX}) it is possible to define a percolation
temperature for these clusters. 

Moreover, the right-hand side expression of
Eq.(\ref{partition_function}) is well defined even for non-integer
values of $s$.  
In these cases it does not define a Hamiltonian model, but it still
defines a percolation model.  
In particular, the $s=1/2$ case for $X>0$ is the {\em frustrated percolation}
model \cite{FFdCC}.

For $X=0$, in the general case in which we consider Potts variables
$\tau_i$ with a generic number $q$ of states instead of an even ($2s$)
number of states in Eqs.(\ref{hamiltonianX},\ref{W_q}), it is possible
to show \cite{Wu} that the percolation temperature $T_p(q,X=0)$ of the
above defined clusters coincides with the ferromagnetic $q$-state Potts
critical temperature $T_c(q)$.  
In particular, in 2D it is possible to prove \cite{Wu} the relation
\begin{equation}
\frac{k_B T_c(q)}{qJ}=\frac{1}{\ln(1+\sqrt{q})} \ .
\label{TcPotts}
\end{equation}

For $X=0.5$ the same kind of relation between the percolation
temperature $T_p(q,X=0.5)$ and the Potts transition temperature 
for $q=2s$ in 2D has been extended \cite{PSG}.  
Furthermore, it was shown numerically that it is possible to generalize the
Eq.(\ref{TcPotts}) using a fitting parameter $a(X)$ \cite{PSG}. 
The resulting relation is
\begin{equation}
\frac{k_B T_p(q,X)}{a(X)qJ}=\frac{1}{\ln\left[1+\sqrt{a(X)q}\right]} 
\label{TpX=0.5}
\end{equation}
with $a(X=0.5)=0.800\pm 0.003$ \cite{PSG}.
Note that $T_p(q,X=0.5)$ represents a percolation temperature for any
$q\geq 0$ and also a Potts transition temperature for even integer values
$q=2s$.
Furthermore, for $0<X<1$ and $q=2s$ the Eq.(\ref{TcPotts}) for
$T_c(2s)$ gives by definition the Griffiths temperature of the
disordered model.

\section{Numerical results in $2D$ for $X=1$}

To study the phase diagram of the model with $X=1$ (Potts FF) we have
simulated it for $s=2$, 7, 20, 50 on square FF lattices with periodic
boundary conditions and linear size $L$ from 10 to 80.
At low temperature all the Potts variables tend to order
ferromagnetically wherever the Ising spins satisfy the
ferro/antiferromagnetic interactions
(i.e. $\delta_{\epsilon_{i,j}S_i,S_j}=1$).  
In particular, at $T=0$ the system in 2D has a second-order phase
transition of the Ising spins, like in the FF Ising model \cite{Villain}.
We will show that the interplay between Potts variables and Ising spins
affects the phase diagram at finite temperature. 
As in Ref.~\cite{PSG}, to study the finite temperature range we can use
an efficient cluster dynamics with an annealing procedure. 
We define a MC step as an update of the whole system.
At each temperature we average the data over $10^4$ MC steps, discarding
the first $5 \times 10^3$ MC steps.

In our systematic analysis we calculated for each $s$ the Binder's
parameter \cite{Binder} for the energy density $E$ defined as
\begin{equation}
V=1-\frac{\langle E^4 \rangle}{3\langle E^2 \rangle^2} 
\label{V}
\end{equation}
(angular brackets denote the thermal average). 
This quantity allows to distinguish between first-order and second-order
phase transitions. 
Indeed, in a second-order phase transition for $L\rightarrow \infty$ it
is $V=2/3$ for all temperatures, while in a first-order phase transition
$V$ has a well pronounced minimum near the transition temperature.
The thermodynamic of the model is studied by means of the Potts order
parameter 
\begin{equation}
M=\frac{s~\mbox{max}_i(M_i)-1}{s-1}
\label{M}
\end{equation}
(where $i=1, \dots s$, $M_i$ is the density of Potts spins in the
$i$-th state), the susceptibility
\begin{equation}
\chi=\frac{1}{k_B T}\frac{\langle M^2 \rangle - \langle M \rangle^2}{N}
\label{chi}
\end{equation}
(where $N$ is the total number of Potts spins) and the specific heat
\begin{equation}
C_H=\frac{1}{k_B T^2}\frac{\langle E^2 \rangle - \langle E \rangle^2}{N}.
\label{cv1}
\end{equation}

To study the Fortuin-Kasteleyn percolation 
we calculated the percolation probability per spin
$P=1-m_1$ and the mean cluster size $S=m_2$
(where $m_n=\sum_k k^n n_k$
is the $n$-th moment of the distribution of density $n_k$ of clusters with size
$k$).

\subsection{The second-order transition for $s=2$}

\begin{figure}
\mbox{ \epsfxsize=8cm \epsffile{ 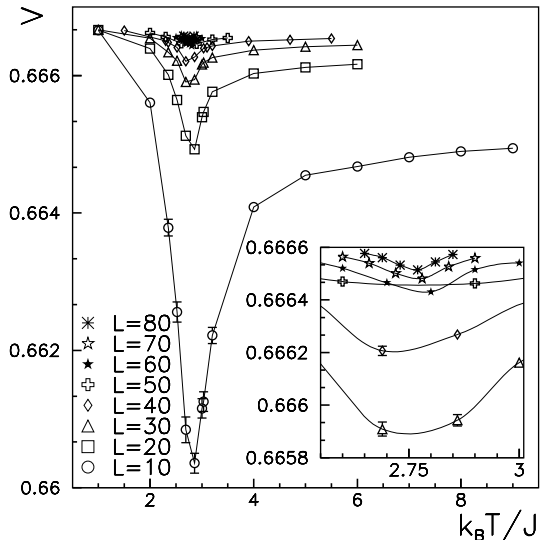 } }
\caption{ $X=1$ and $s=2$: Binder's parameter $V$
vs. $T$ for the lattice sizes $L$ listed in the figure. Inset: Enlarged view.
Where not shown, errors are smaller than symbols size.
Lines are only guides for the eyes. For increasing $L$,
$V$ goes to $2/3$ for all temperatures showing a second-order phase transition.
}
\label{V4}
\end{figure}

For $X=1$ and $s=2$ the Binder's parameter $V$ goes to the constant
value 2/3 for all the temperatures as $L$ increases (see Fig.~\ref{V4}) 
revealing a second-order phase transition. 
The transition temperature $T_s$ in the thermodynamic limit can be estimated,
together with all the critical exponents, using the standard 
scaling analysis \cite{Bin88} for second-order phase transition, 
where by definition of critical exponents $\nu$, $\beta$, $\gamma$ and 
$\alpha$ is 
$\xi\sim|T-T_s|^{-\nu} $
being $\xi$ the correlation length,
$ M\sim|T-T_s|^{\beta}\sim \xi^{-\beta/\nu} $,
$ \chi\sim|T-T_s|^{-\gamma}\sim \xi^{\gamma/\nu} $,
$ C_H\sim|T-T_s|^{-\alpha}\sim \xi^{\alpha/\nu}  $,
for which we expect 

\begin{equation}
M\sim L^{-\beta/\nu} f_M((T-T_s)L^{1/\nu})     ~,
\label{scaling_M}
\end{equation}
(and analogous scaling laws for $\chi$ and $C_H$)
where $f_M(x)$ is an universal function of the 
dimensionless variable $x$.
The values at which the scaling laws
are verified give the estimates of critical exponents and of $T_s$.

\begin{figure}
\mbox{ \epsfxsize=8cm \epsffile{ 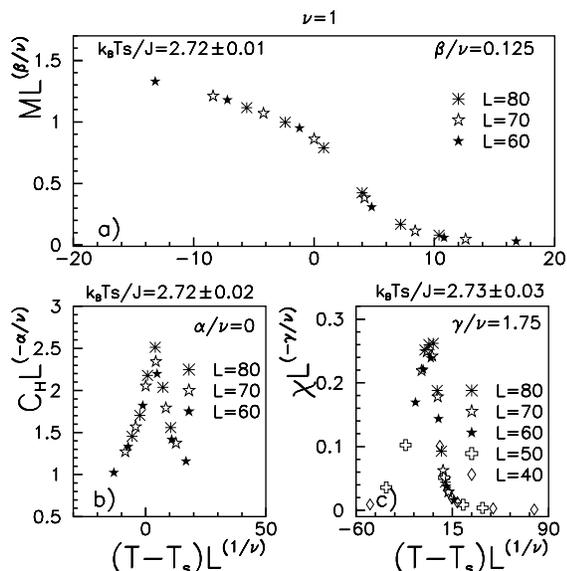 } }
\caption{ $X=1$ and $s=2$: Collapse of $M$, $C_H$ and $\chi$ data in a),
b) and c) panel, respectively, 
for the Ising
critical exponents ($\nu$, $\beta$, $\alpha$, $\gamma$). Each collapse
gives an independent estimate of the critical temperature $T_s$
(indicated in each panel).
}
\label{sca_s2}
\end{figure}

In Figs. \ref{sca_s2} we show the large-sizes data collapses 
using the set of Ising critical exponents
\cite{Stanley}
and leaving only $T_s$ as free parameter, giving an estimate 
$k_B T_s/J=2.73\pm 0.03$.

\begin{figure}
\mbox{ \epsfxsize=8cm \epsffile{ 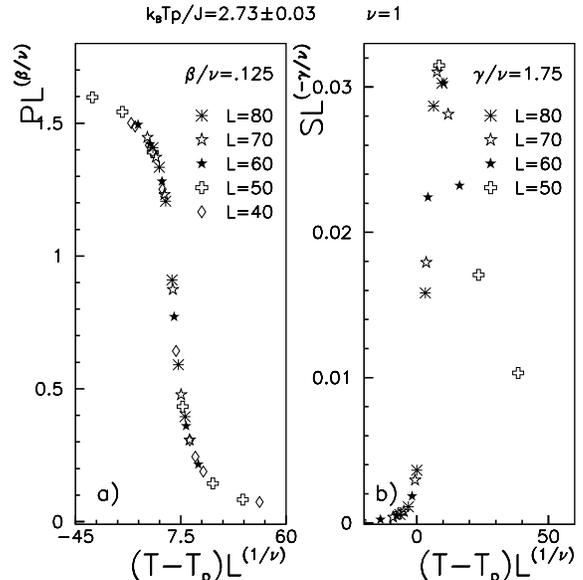 } }
\caption{ $X=1$ and $s=2$: Collapse of $P$ and $S$ data in panel a) and
b), respectively, for the Ising
critical exponents. Each collapse gives an estimate of the percolation
temperature $k_B T_p/J=2.73\pm 0.03$. 
}
\label{sca_P}
\end{figure}

The study of the percolation quantities shows a 
smooth behavior of $P$ and a cusp in $S$ increasing with $L$. Therefore 
one can make the ansatz that the percolation transition is of
second-order. Let us  define a Fortuin-Kasteleyn percolation temperature
$T_p$  and
a set of percolation critical exponents 
$\nu_p$, $\beta_p$ and $\gamma_p$ by means of the relations
$ \xi_p\sim|T-T_p|^{-\nu_p} $
where $\xi_p$ is the connectedness length of the clusters
(i.e. the typical linear cluster size) and
$ P\sim |T-T_p|^{\beta_p}\sim \xi_p^{-\beta_p/\nu_p} $,
$ S\sim|T-T_p|^{-\gamma_p}\sim \xi_p^{\gamma_p/\nu_p} $.
Applying the 
standard scaling analysis for percolation \cite{Stauffer_Aha},
one obtains in a consistent way the 
results summarized in Fig. \ref{sca_P}
using the
corresponding thermodynamic critical exponents for the 2D Ising model.
The numerical estimate for the percolation temperature is $k_B T_p/J=
2.73 \pm 0.03$ coincident with the estimates of $T_s$.
Therefore the percolation transition and the Potts
transition occur at the same temperature and with the same set of
critical exponents, i.e. they coincide.

\subsection{The first-order transition for $s=7$, $20$ and $50$}

For $X=1$ and  $s=7$, 20 and 50 we have considered systems with
$L$ from 10 to 50 lattice steps with periodic boundary conditions.
On the base of the mean field results \cite{dLP} for the Potts FF model and the
knowledge of the Potts model \cite{Wu}, we expect that the order of transition
changes for $s>4$. In fact,
the thermodynamic order parameter $M$ becomes more and 
more discontinuous as $s$ increases. At the same time the percolation 
order parameter $P$ develops a
 better and better pronounced discontinuity.

\begin{figure}
\mbox{ \epsfxsize=8cm \epsffile{ 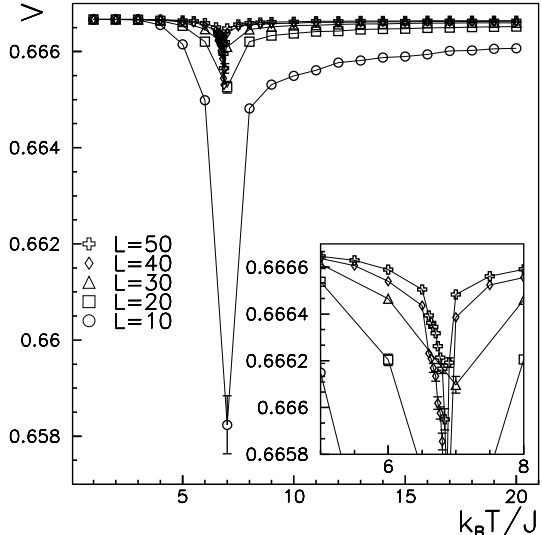 } }
\caption{ $X=1$ and $s=7$: Binder's parameter $V$
as in Fig.2. $V$ has a well defined minimum for any
size, showing a first-order phase transition.
}
\label{V7}
\end{figure}

In particular, the study of the Binder's parameter $V$ reveals that the 
model for $s=7$, 20 and 50 has a first-order phase transition, since, 
for each considered lattice size $L$ between 10 and 50, $V$ has a non 
vanishing minimum, as shown in Fig. \ref{V7} for $s=7$.
In this cases the estimates of infinite size transition temperatures
$T_s(s)$ and $T_p(s)$, 
for the thermodynamic and the percolation transition, respectively, can
be done through  
the relation \cite{Bin88}
\begin{equation}
T_{\mbox{max}}(L)-T_{\mbox{max}}(\infty)\sim L^{-D}
\label{Tp(L)}
\end{equation}
where $D=2$ is the Euclidean dimension,
$T_{\mbox{max}}(L)$ is the finite-size temperature of the maximum
of $C_H$ and $S$, respectively, and
$T_{\mbox{max}}(\infty)$ is the transition temperature in the
thermodynamic limit. 
Therefore $T_s(s)$ and $T_p(s)$ can be evaluated by linear fits on a
Log-Log scale with one free parameter. 
The results are summarized in Tab.~I.
For any $s$, $T_s(s)$ and $T_p(s)$ are consistent within the numerical
error.

\subsection{Phase diagram}

\begin{figure}
\mbox{ \epsfxsize=8cm \epsffile{ 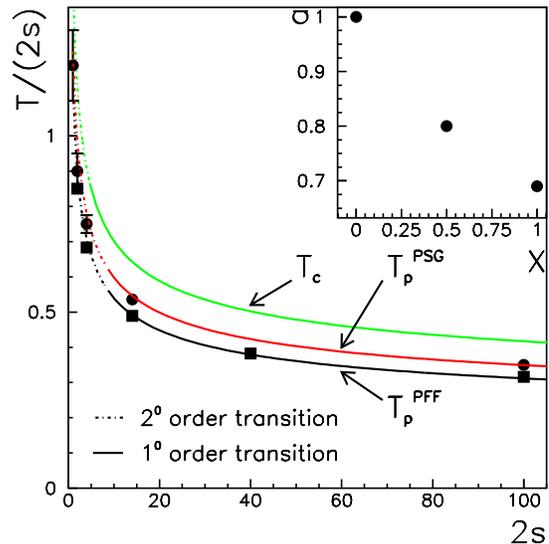 } }
\caption{Numerical phase diagram in $2D$ for $X=0$ (Potts model: $T_c$),
$X=0.5$ (Potts SG model: $T_p^{PSG}$) and $X=1$ (Potts FF model: $T_p^{PFF}$). 
The data are fitted with
Eq.(\protect\ref{TpX=0.5}) with fit parameter $a(X)$ shown in the inset. 
Errors are smaller than symbols size.
}
\label{phase_d}
\end{figure}

The numerical results in Secs.~IV.A and IV.B give rise to the phase
diagram for $X=1$. 
It is qualitatively similar to the phase diagram for $X=0.5$ \cite{PSG}. 
In Fig.\ref{phase_d} we compare both of them with the phase diagrams for
$X=0$.

For $X=0$ (Potts model with $2s$-states variables) for any
integer $s$ there is a paramagnetic/ferromagnetic  phase transition at finite
temperature $T_c(s)$. 
The transition is of second-order for $2s\leq 4$ and of first-order for
$2s>4$  \cite{Wu}. 
$T_c(s)$ is given by Eq.(\ref{TcPotts}) with $q=2s$. 
It coincides with the Fortuin-Kasteleyn percolation temperature. 
The percolation temperature is defined even for non-integer values of
$s$. 

For $X=0.5$ (Potts SG model with frustration induced by disorder and
with $2s$-states variables) there are two phase transitions for integer
values of $s$. 
The lower transition is a SG transition.
It is  supposed at $T_{SG}=0$ for any $s$ in 2D and at $T_{SG}(s)>0$ in
higher dimensions as in the Ising SG \cite{KawashimaAoki}. 
The high-temperature transition is a $s$-states Potts
ferromagnetic transition occurring at $T_p(s,X)$ given by
Eq.(\ref{TpX=0.5}) with $q=2s$ and $a(X=0.5)=0.800\pm 0.003$
\cite{PSG}. 
It is a second-order transition for $s\leq 4$ and a first-order
transition for $s>4$. 
$T_p(s,X)$ corresponds also to the Fortuin-Kasteleyn percolation
temperature, that is defined even for non-integer $s$ \cite{PSG}. 
This model  is disordered and its Griffiths 
temperature is by definition the transition temperature of the
$2s$-states Potts model $T_c(s)$, given by the Eq.(\ref{TcPotts}) with
$q=2s$. 
It is possible to see that $T_c(s)$ is numerically consistent with the
onset of non-exponential relaxations at $T=T^*(s)$ for $s=2$ in 2D
\cite{breve} and for $s=1$ (Ising SG) in 2D and 3D
\cite{Ogielski,McMillan}. 

For $X=1$ (Potts FF model with frustration and no disorder and
with $2s$-states variables), considered here, there are two
transitions for integer values of $s$, as well as for $X=0.5$.
The lower transition is at $T_{FF}=0$ in 2D and at $T_{FF}(s)>0$ in
higher dimensions, as for the FF Ising model ($s=1$)
\cite{Villain,FF3D}. 
As seen in this section, in this case the upper transition is a
$s$-states Potts ferromagnetic transition at $T_p(s,X=1)$. 
As shown in Fig.~\ref{phase_d}, Eq.(\ref{TpX=0.5}) describes
well $T_p(s,X=1)$, using a fit parameter $a(X=1)=0.690\pm 0.003$. 
Furthermore we have shown that it coincides with a Fortuin-Kasteleyn
percolation transition.
As consequence the Fortuin-Kasteleyn clusters represent the
regions of correlated Potts variables.
This means that the cluster's characteristic linear size is equal to the
correlation length of the Potts variables.
Analysis for $s=1/2$ (frustrated percolation) and $s=1$ (FF Ising model) 
in 2D and 3D are given in Ref.\cite{FFdCC}. 

The main difference between $X=1$ and $X=0.5$ cases is that the
Griffiths temperature is defined in the latter (disordered model), but
not in the former (for the lack of disorder). 
The important consequence of this fact is that for $X=1$ the dynamic
anomalies are present only below $T_p(s,X=1)$ \cite{X=1},
while for $X=0.5$ they are present also above $T_p(s,X=0.5)$ and
below $T_c(s)$ \cite{X=0.5}.

Note that all the Potts-percolation transition temperatures for $X=0$,
0.5, 1 can be described by the same form in Eq.(\ref{TpX=0.5}) as
function of $s$ with a $X$-dependent parameter $a(X)$. 
This parameter has a regular behavior as function of $X$ (see the
inset in Fig.\ref{phase_d}).

\begin{figure}
\mbox{ \epsfxsize=8cm \epsffile{ 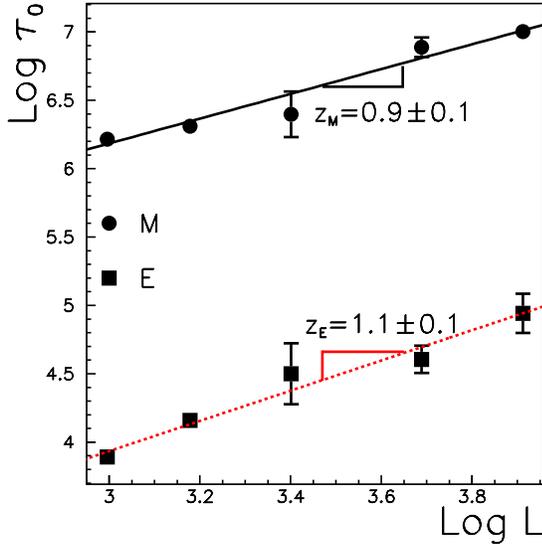 } }
\caption{ $X=1$ and $s=2$:
The logarithm of correlation time $\tau_0$ (see text) at the finite-size
transition temperature $T_p(L)$ \protect\cite{nota_Tp} as function of
the logarithm of size $L$, for the correlation function of the Potts order
parameter $M$ (circles) and of the energy density $E$ (squares), 
for $L=20$, 24, 30, 40, 50.
The slopes give the exponents $z_M$ and $z_E$.
}
\label{fig_z}
\end{figure}

\section{Dynamic critical exponent}

As shown in Sec.~IV.A, at $T_p(s=2,X=1)$ the Potts variables have a
second-order phase transition.
Therefore their correlation length diverges and their dynamics slows down. 
A measure of the slowing down for any observables $A$ is given by the
dynamic critical exponent $z$ defined by $\tau_A(T_p(L),L)\sim L^z$.
Here, omitting for sake of simplicity the dependence on $s$ and $X$, 
$T_p(L)$ is the transition temperature for the system with finite size 
$L$ \cite{nota_Tp} and  $\tau_A(T,L)$ is the correlation time at 
temperature $T$ and size $L$ associated to the correlation function
for $A$
\begin{equation}
f_A(t,T)=\frac{\langle A(t,T) A(0,T)\rangle - \langle A(T) \rangle^2}
{\langle A(0,T)^2 \rangle - \langle A(T) \rangle^2},
\end{equation}
where $t$ is the time.
More then one definition of $\tau_A$ is possible for any $A$, but all of
them, even if numerically different, have the same qualitatively
behavior \cite{Ogielski}.
In particular, facing the difficulty that the greater size the greater 
$\tau_A(T_p(L),L)$, we define $\tau_M$ and $\tau_E$ for  Potts order
parameter $M$ and for energy density $E$ as the time $\tau_0$ (in unit
of MC steps) at which $f_M(\tau_0,T_p(L))=0.4$ and
$f_E(\tau_0,T_p(L))=0.3$, respectively. 
The data for sizes $L=20$, 24, 30, 40, 50 are shown in Fig.\ref{fig_z}
and the dynamic critical exponents are estimated by linear fits on
Log-Log plot as $z_M=0.9\pm 0.1$ and $z_E=1.1\pm 0.1$, respectively
\cite{nota_z}. 

\begin{figure}
\mbox{ \epsfxsize=8cm \epsffile{ 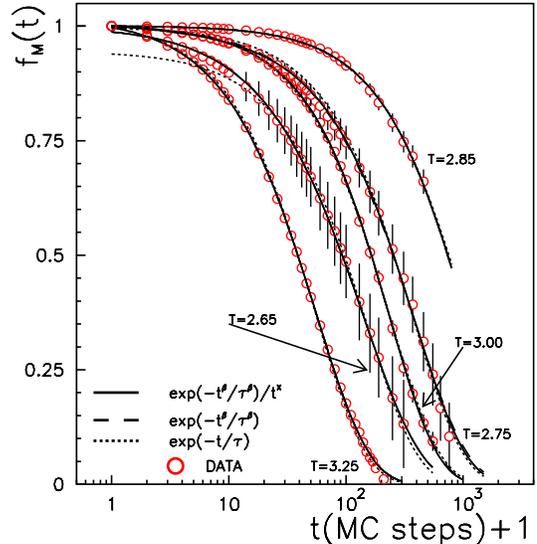 } }
\caption{ $X=1$ and $s=2$:
Correlation function of the Potts order parameter $M$
in the thermodynamic limit (see text). For clarity we show only some of the recorded 
data for the simulated temperatures. Symbols are results of simulations, solid lines
are fits with the form in Eq.(\protect\ref{stretch}),
dash lines (on this scale indistinguishable from solid lines) 
with stretched exponential form 
and dotted lines with exponential form. 
Where not shown, the errors are smaller that the
symbols size. Temperature is measured in $J/k_B$.
}
\label{corr_M}
\end{figure}

\begin{figure}
\mbox{ \epsfxsize=8cm \epsffile{ 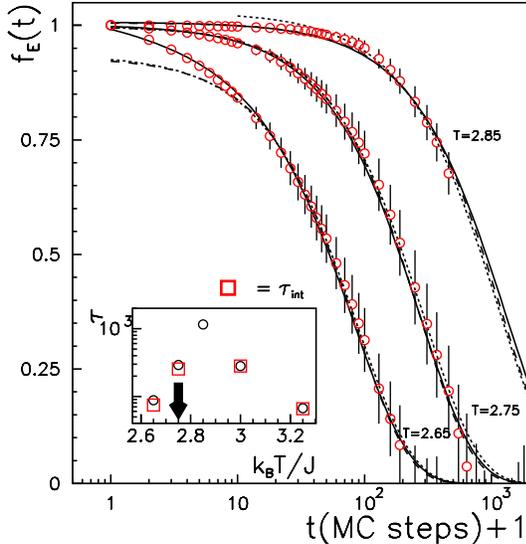 } }
\caption{ $X=1$ and $s=2$: 
Correlation function of the energy density $E$
in the thermodynamic limit as in Fig.8.
Inset: The correlation times estimated by exponential fit (circles) and by
Eq.~(\protect\ref{int}) (squares). 
The arrow shows the numerical estimate
of $T_p$.
}
\label{corr_E}
\end{figure}

To study the behavior of $f_M$ and $f_E$ in the thermodynamic limit 
we have  extrapolated the data in the infinite size limit following the
procedure suggested in Ref.\cite{Campbell_Bernardi}. 
It consists in  plotting at any $t$ the generic $f_A(t,T,L)$ for finite
size $L$ versus $1/L$ and in extrapolating for $1/L\rightarrow 0$. 
We consider the temperature range $2.65<k_B T/J<3.25$ \cite{nota_corr}.
The results are shown in Figs. \ref{corr_M} and \ref{corr_E}. 
To check the form of $f_M$ and $f_E$, we have fitted the data
with three different plausible functions:
$a)$ Simple exponential $f_0 \cdot \exp(-t/\tau)$; 
$b)$ Kolraush-Williams-Watts stretched exponential $f_0 \cdot
\exp[-(t/\tau)^\beta]$;
$c)$ Ogielski form 
\begin{equation}
f(t,T)=f_0 \frac{e^{-(t/\tau)^\beta}}{t^x} 
\label{stretch}
\end{equation}
that is a combination of a form $b)$ with a power law. 
In the previous functions, $f_0$, $\tau$, $x$ and $\beta$ are
$T$-dependent parameters. 
Note that for the form $c)$ it is possible to estimate $f_0$ and
$x$ separately from $\beta$ and $\tau$, since the first two describe
the short-time behavior while the second two the long-time regime.

\begin{figure}
\mbox{ \epsfxsize=8cm \epsffile{ 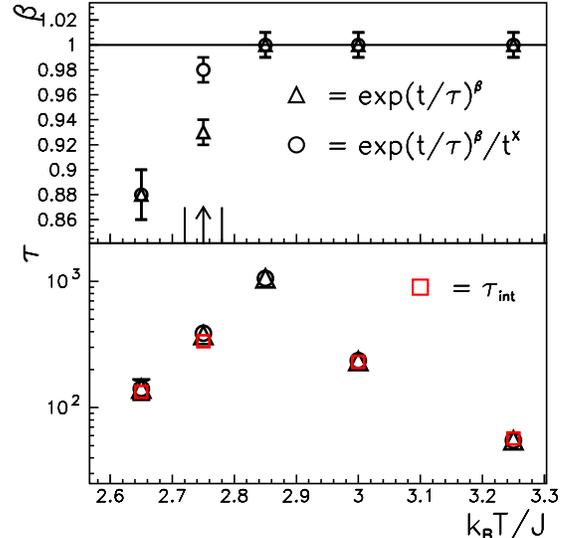 } }
\caption{ $X=1$ and $s=2$: Fit parameters for $f_M$ data with 
form in Eq.~(\protect\ref{stretch})
 (circles) and for stretched exponential form 
(triangles). In the lower panel we show also the integral correlation times 
$\tau_{int}$ estimated 
with the Eq.(\protect\ref{int}) (squares). 
The arrow and the vertical lines show 
$T_p$ and the numerical indetermination on it, respectively.
}
\label{par_M}
\end{figure}

The Ogielski form turns out to describe very well $f_M$ and $f_E$
as shown in Figs. \ref{corr_M} and \ref{corr_E}.
The fitting parameters are presented in Fig.\ref{corr_E} and \ref{par_M}.
The forms $b)$ and $c)$ give always compatible estimates of $\beta$
exponent, consistently with the very low values of $x$ (that is
approximately the slope of the function at $t=0$ in Figs.\ref{corr_M}
and \ref{corr_E}).  

From Fig.~\ref{par_M} for the parameter $\beta$ of $f_M$, we see that it
is $\beta=1$ above the Potts transition at $k_BT_p/J=2.73\pm 0.03$ and
$\beta<1$ below $T_p$. 
Therefore there is a dynamic transition between a high-temperature
exponential behavior to a low-temperature stretched
exponential behavior of $f_M$.
This result shows the presence of a complex dynamics below $T_p$,
consistently with the analysis of non-linear susceptibility correlation
function of the Ising spins \cite{breve}. 
It is important to note that, while the Potts variables have a
transition at $T_p$, the Ising spins have no transition at $T_p$ and, in
principle, no dynamical anomalies are expected for them. 

On the other hand, the data for $f_E$ show that the long-time behavior
is well described by an exponential function. 
In particular, the form $c)$ with $\beta=1$ and a non-zero $x$ fits very
well also the data for the lowest temperature considered here.
Note that, in principle, one can expect a dynamical anomaly even for the
energy density $E$, since it depends explicitly on the critical Potts
variables. 

In the fitting forms used, the correlation time $\tau$ is defined as
a fitting parameter. 
Another possible definition is the following 
(integral correlation time):
\begin{equation}
\tau_{int,A}(T)=
\lim_{t_{max}\rightarrow \infty}\frac{1}{2}+\sum_{t=0}^{t_{max}}f_A(t,T)
\ .
\label{int}
\end{equation}
Due to the divergence of the correlation time, the definition in
Eq.(\ref{int}) does not converge near the transition temperature.
Therefore it is not shown in Figs.~\ref{corr_E} and \ref{par_M} for the
temperature with the largest $\tau$.  
However, where it converges, all the different estimates of the correlation
time  are numerically consistent as show in Fig.\ref{corr_E} and \ref{par_M}.
Note that, even if the data are extrapolated to the thermodynamic limit,
finite size effects are still present. 
Indeed, the critical temperature estimated from $\tau$ measurements,
within our temperature mesh, is between $2.75 J/k_B$ and $2.85 J/k_B$,
that is at variance with the estimate of $T_p$.

\section{Summary and Conclusions}

We have studied a schematic model for glasses where frustrated orientational  
degrees of freedom, associated to $2s$-states Potts variables $\tau_i$, induce
complex dynamics. 
In real systems the frustration can be associated, for example, to
steric hindrance of non spherical molecules in structural glasses.
Each $\tau_i$ has an absolute value ($\sigma_i=1, \dots , s$) and a sign
($S_i=\pm 1$).  
The frustration is over the signs and can be {\em with} disorder or
{\em without} disorder, depending on a model parameter $X$.
For $0<X<1$ there is disorder, while for $X=1$ the model is (fully)
frustrated without disorder. 
For $X=0$ there is no frustration, and the model recovers the
$2s$-states ferromagnetic Potts model.

For $0<X\leq 1$ the model has two thermodynamic transitions.
The high-temperature transition at $T_p(s,X)$ for the model with $2s$
states is in the universality class of the $s$-states ferromagnetic
Potts model. 
Therefore the fluctuations ($\chi$ and $C_H$) of the orientational
degrees of freedom diverge at $T_p(s,X)$ (for $s\leq 4$), as well as the
correlation times of quantities depending on them (like the Potts order
parameter $M$ or the energy density $E$).
For them we estimated the dynamic critical exponents $z_M$ and $z_E$. 
The low-temperature transition is a spin glass transition for $X=0.5$,
or a fully frustrated transition for $X=1$, and marks the ordering
transition of  the signs $S_i$ (Ising variables).

The diverging fluctuations at upper and lower transition temperatures 
are expected to be experimentally observable only using specific probes
that couple with them. 
Examples of such probes could be those associated to dielectric
measurements in supercooled-liquids and plastic glassy crystals or to
electron spin resonance spectroscopy measurements \cite{glass,Andreozzi}.

For $X=1$ and $s=2$ the model shows a complex dynamics in correspondence
of $T_p$.  
In particular, the correlation function for $M$ and the correlation
function for the signs $S_i$ change their behaviors. 
They have an exponential behavior above $T_p$ and a non-exponential
behavior below $T_p$.
This behavior is expected at least for any $s\leq 4$, because it is
associated to the free energy singularity occurring at the second-order
transition of the Potts variables. 
Therefore the complex dynamics corresponds to a {\em real}
thermodynamic transition.

For $X=0.5$ and $s=2$ the onset of this dynamic anomaly is shifted to a
higher temperature, {\em above} $T_p$. 
It occurs in correspondence of the Griffiths essential singularity of
the free energy.
This singularity is not defined in the $X=1$ case.
For zero external field it goes to the transition temperature of the
ferromagnetic $2s$-states Potts model $T_c(2s)$ and it vanishes.
Therefore the onset of complex behavior does not correspond to a {\em
real} thermodynamic transition.

We have shown that the Potts transition at $T_p(s,X)$ for any $s$ and
$X$ considered here, coincides with a percolation transition.
It is not worthless to note that the dynamic transition at $T_p(s,X=1)$ 
persists also for $s=1$ and $s=1/2$ \cite{FFdCC}. 
In these cases $T_p(s,X=1)$ does not correspond to a thermodynamic
transition, but only to a percolation transition.
Therefore for $s<2$ the dynamic anomaly is no more related to a
thermodynamic transition, but to a percolation transition in the real
space \cite{FFdCC}.
This result could be in some relation with experimental results on
microemulsions \cite{Feldman}. 

Finally, we have shown that it is possible to generalize the exact
relation for the ferromagnetic Potts transition temperature $T_c(s)$ in
2D to a transition temperature $T_p(s,X)$ for any $0\leq X \leq 1$,
using a fitting parameter $a(X)$. 
In particular, $a(X)$ acts like a renomarlization factor for the number of
states of the model and it is $a(X=0)=1$ (ferromagnetic case) and $a(X)$
decreasing regularly with increasing $X$.  

\section*{Acknowledgments}

I am grateful to A.~Coniglio for many stimulating
observations. I would like to thank G.~Parisi, S.~Franz and
Y.~Feldman for interesting discussion and  L.~Amaral and A.~Scala for a
critical reading of the manuscript. 
Partial support was given by
the European TMR Network-Fractals c.n.FMRXCT980183.

\begin{table}
\begin{tabular}{c c c}
$s$ & $T_s$          & $T_p$ \\
\hline
7   & $6.87\pm 0.04$ & $6.85 \pm 0.06$\\
20  & $15.3\pm 0.1$  & $15.3 \pm 0.1$\\
50  & $31.7\pm 0.1$  & $31.5 \pm 0.1$
\label{tabII}
\end{tabular}
\caption{$X=1$ and $s=7$, 20, 50: Numerical estimates of thermodynamic transition 
temperature
$T_s(s)$ and percolation transition temperature $T_p(s)$. For any $s$
they are consistent within the numerical error.
}
\end{table}

\end{multicols}


\begin{references}
\bibitem{nota_CK}
In the Coniglio-Klein approach \cite{FK-CK} a bond can be placed with
probability $p$ given by the expression after Eq.(\ref{W_q}) 
only between nearest neighbor
sites $i$ and $j$
with $\delta_{\epsilon_{i,j}\tau_i,\tau_j}=1$
(e.g. the couples $a$, $c$,
$e$ and $g$ in Fig.~\ref{lattice}) and there is no need to define the
frustrated loops, because they are automatically excluded. 
However this procedure is defined only for integer values of $s$.
\bibitem{nota_Tp}
$T_p(L)$ here is defined as the temperature at which the largest
correlation time occurs.
\bibitem{nota_z}
Each data in Fig.\protect\ref{fig_z} is the average on 16 independent
simulations with a standard local spin-flip dynamics \cite{Bin88}
with an annealing method starting from $k_B T/J=3.25$,
with $10^4$ equilibration MC steps at each temperature, averaging the
data over $10^4$ MC steps for $L=30$, 40 and over $2\times 10^4$ MC steps
for $L=20$, 24, 50.
\bibitem{nota_corr}
For each $T$ and $L=20$, 24, 30, 40, 50 the 
data are obtained annealing the system 
with $10^4$ equilibration MC steps and averaging over more then $10^6$
MC steps.
\end{references}
\end{document}